# Towards a Systems Engineering Essence[1]

*Anatoly Levenchuk*
*ailev@asmp.msk.su*

**Abstract**

SEMAT/OMG Essence provides a powerful Language and a Kernel for describing software development processes. How can it be tweaked to apply it to systems engineering methods description? We must harmonize Essence and various systems engineering standards in order to provide a more formal system approach to obtaining a Systems Engineering Essence. In this paper, an approach of using Essence for systems engineering is presented. In this approach we partly modified a Kernel only within engineering solution area of concerns and completely preserved Language as an excellent situational method engineering foundation.

**Introduction**

Systems engineering is highly diversified in regard to Systems-of-Interest types and methods utilized in their definition and realization. Moreover, the discipline is rapidly changing just now by adopting contemporary model-based methodologies. By definition systems engineering has blurry boundaries with control systems (cyber-physical system) engineering, software engineering and even enterprise engineering that is also considered to be in the domain of the systems engineering discipline. One can question: "What is the systems engineering knowledge that will persist? "

One of the source of an explicitly declared persisting and spanning multiple engineering domains systems engineering knowledge is standardization. New and promising concepts of a systems engineering body of knowledge are being discovered and discussed, then published for general engineering community usage. This includes publications approved by professional bodies including Handbooks (SEH, 2014), classical Body of Knowledge (BKCASE, 2014) and more narrow subdomain standards that cover particular aspects of systems engineering practices. *Note: The appendix provides a brief description of the relevant standards and publications that are referred to in this text.*

Contemporary systems engineering has benefited from software engineering standards but with a lag of 10-15 years. SysML (OMG SysML, 2012) of systems engineering is based upon UML (OMG UML, 2011) of software engineering but appears later. Agile in systems engineering processes started in last few years. OMG SPEM 1.0 was only about software, SPEM 2.0 now covers software and systems engineering (OMG SPEM, 2008).

---

[1] This work was recommended as an *INCOSE Russian chapter product* at 99th INCOSE Russian chapter meeting, 28-Jan-2015.

Systems engineering differs from Engineering of systems in a special way of embracing a system approach. Systems engineering and systems engineering management are all about the routine and ubiquitous usage of system thinking. Engineering of systems presumes the naming of each and every object as a "system" but does not explicitly utilize system thinking (Hitchkins, 2010).

If we want to develop the true conceptual core of systems engineering (A Systems Engineering Essence), we need to harmonize multiple software and systems engineering standards that are based on a system perspective that includes system thinking and system acting (Lawson, 2010 and Lawson, 2015). We believe that it is possible to tweak the software-based OMG Essence (OMG Essence, 2014) for usage in systems engineering as it were accomplished with UML-SysML, SPEM 1.0-SPEM 2.0, etc.. We need to create a Systems Engineering Essence based upon the OMG software engineering Essence.

The joint project of the INCOSE Russian chapter and SEMAT in 2013-1014 was devoted to research concerning possible ways of adapting Essence for systems engineering. The INCOSE Russian chapter is working towards harmonizing the joint usage of system thinking in OMG Essence with the international standards ISO/IEC/IEEE 15288, ISO/IEC/IEEE 42010, IEC 81346 and ISO 15926. Our aim is to obtain a solid, due to standards as a foundation, system thinking and acting framework for systems engineering.

## How to describe systems engineering

### Terminology and ontology

In relation to terminology of systems, software and enterprise engineering, it is important to understand the difference between speech communities and semantic communities (OMG SBVR, 2013).

Speech communities share a common language and vocabulary for it: natural language (English, Russian, etc.) or terms from some kind of standard (OMG Essence, ISO 15926, etc.).

Semantic communities share a unified understanding of some interrelated concepts set (discipline). E.g. the semantic community understands "car" as a concept without confusing it with "bicycle" or "ship" whether it be called with a term "car", "автомобиль" or "motor vehicle".

Systems engineers of the world form a common semantic community, but each of them typically belonging to a particular speech community. These systems engineering speech communities each utilize their own systems engineering terminology and normative vocabulary. Thus, we need not rely too much upon the specific terms and designations but should always think about the concepts behind them.

It is very tricky. For example, the term "Function" can refer to at least 5 different concepts (Renssen, 2005):

    1. An occurrence (activity, process or event).

    2. A totality in a particular role or designed or made for an intended role.

3. A role of a totality (usually of a physically qualified thing) played in an occurrence.

4. A correlation, usually as a physical coupling between aspects (if the magnitude of one aspect changes, then the magnitude of the other necessarily changes as well, in a particular way).

5. A mathematical relation between numeric objects, which specifies a mapping.

System function: what is it? We need a solid systems engineering ontology to have an answer to such a basic question. The Essence Kernel provides an ontology as a "common ground" for software engineering. "An ontology is a formal specification of a shared conceptualization" (Gruber, T., 1995). Standardization helps in claiming that a system engineering ontology is shared thus not only providing a "ground" for discussing of systems engineering method, but a "common ground".

(Sowa, J., 2012) has indicated that we need an underspecified ontology to be useful, but it should not be too general to preserve the value of using it in a particular domain, in our case the systems engineering domain. We need to find the "right level" of generalization in order to capture the essence of the systems engineering domain.

The Essence Kernel has established a simple approach to finding such an ontology for software engineering: first, they separate alpha (concept from discipline) and work product (realization of the concept in particular project context) and present it as the Essence Language, a meta-meta-model of an engineering project. Second, they attempted to find all concepts that are used in about a 250 software process methodologies and expressed the most ubiquitous of them in that Language. The result is that the Essence kernel provides a meta-model of an engineering project (Jacobson, 2015).

We are tweaking and amending the Essence kernel with concepts from systems and software engineering standards (ISO/IEC/IEEE 15288, ISO/IEC/IEEE 42010, IEC 81346 and ISO 15926 first of all) to provide a systems engineering ontology ("common ground"). Our research is at a preliminary stage but it is very promising.

### Theory of systems engineering

The education of systems engineers is becoming more similar to education in the humanities than in the sciences. Also engineering is about heuristics, not theories (Koen, 2003). But we need more logic in our knowledge about method. We need some kind of formalism for promoting systems engineering method discussions. We need logical (not necessary Aristotelian, like not every geometry is Euclidean) schemas for systems engineering knowledge. Further we need theory, not simply metaphors and style suggestions in the crafting of engineering artifacts.

Contemporary standards unlike many of the textbooks often suggest logic-based schemas that provide rigor for their text. Usually such schemas are provided in diagrammatic form. E.g., OMG Essence have MOF-based diagrams, ISO/IEC/IEEE 42010 utilizes a quasi-UML notation (ISO/IEC/IEEE 42010, 2011), ISO 15926-2 has its own schema notation (ISO 15926-2, 2003).

To teach systems engineering as a science we need to suggest a set of ideal objects that correspond to real world objects and teach students to mentally operate with them. Like physics studies physical bodies that have mass and acceleration properties that do not exist in the real world (a ball and a rocket are not physical bodies, we simply "blend" it with physical body and thus attribute mass and acceleration properties that can be applied for ball and rocket). In a manner similar to chemistry studies of chemical bonds and valences that are ideal objects, we need such ideal objects for systems engineering.

The Essence language suggests the usage of alphas for representing such ideal objects-from-a-discipline and work products for representing objects-from-a-real-world).

We study alphas "requirements" or "team" and train our discipline (as in physics or chemistry) to think about them as we do with any theory. In real life we usually have not seen any "requirements" documents named "The Requirements" and "team" that named "The Team" but only variety of documents or database records provided by the multiple human actors involved. Those actors give them names that have no resemblance to "requirements" and "team". We must somehow "blend" these theoretical concepts (alphas) with real life concepts (work products) in order to apply our theoretical knowledge to real life situations.

The phrase "stakeholders are giving opportunities" is a fact from a theory, which is applicable to multiple situations. In a real life project, this phrase can be stated as: "client X needs system Y that we can developed for them in one month". However, disciplined knowledge about alphas from theory urge us to seek "opportunity" when "stakeholders" are mentioned and bring to our memory "stakeholders" when we deal with "opportunities", whatever object serves as a work product providing evidence for these alphas. Theory (in our case systems engineering theory) is not bound to creativity but preserves some logic in thinking about complex projects (like theory not bounding creativity when we apply physical or chemistry formulas to nature issues).

We cannot be overly precise in our wording. We use schemas but rarely do we have in our speech exact phrases that can be read directly from formal schema. Often we use metonymy (a name associated with a concept) when we identify objects with the name of objects adjacent to them in a schema (while metaphor is not about following relations in a schema but about the similarity of two or even more schemas). Very often metonymy appears in alpha-work product relations. Beware not to mix alphas and work products. E.g. "is architecture approved?" can be about an alpha (engineering solution approved: ideal object) or work product (architectural descriptions approved: a couple of diagrams on a paper). OMG Essence informs us to distinguish one from another, despite metonymy.

The systems engineering semantic community is those people who share a systems engineering ontology, i.e. share formal specification of a systems engineering method conceptualization (e.g. given as a UML-like diagrams, MOF-based etc., that today routinely are utilized in contemporary systems engineering standards).

## Situational method engineering

When we want to speak about method, first we need a Language (upper ontology) for the method in order to share a common understanding. Numerous standards pretend to provide a Language for method definition. E.g. ISO/IEC TR 24774 that was an attempt to develop small "process language" (ISO/IEC/TR 24774, 2010). This Language was used in ISO/IEC/IEEE 12207 and ISO/IEC/IEEE 15288 to describe "processes" structured into a hierarchy of Activities and Tasks. However, there are multiple standards that were suggested by a situational method engineering discipline (ISO/IEC 24744, 2007; OMG SPEM 2.0, 2008). That method engineering is "situational" because no one method works in another situation than that for which it was developed. In situational method engineering standards it is suggested that we can reuse method components (chunks, slices, etc. – there are nuances in different approaches to method decomposition to have reusable method parts). Further, the resulting Language of such method components should give us a convenient means for composition (definition) of a particular method that fit to a particular project (Henderson-Sellers at al., 2010).

There were two generations of situational method engineering standards. Most of them stem from software engineering method description projects. Nevertheless, it appears that such standards are relatively easy also to apply to systems engineering development. OMG Essence is first of second generation standards, first that is based on alpha vs. work product distinction and provides not only "language" but some ontology ("kernel": alphas and activity spaces) for a theory-based (ideal objects-based) defined discipline of «software engineering». Roughly "discipline + work products, tools and activities = practices". Discipline is studied in academic environments, whereas practices are learned in the workplace with given tools and types of work products.

We want to preserve the Language from OMG Essence and modify the Essence Kernel as a base for the definition of systems engineering as a discipline that is composed of multiple and diverse practices that have a common mindset. Modification is needed to Systems Engineering Essence so that it can cover not only software but hardware life cycles as well. We need the most general ontology ("common ground", discipline definition, Kernel, formal diagrammatic domain schema) of engineering that is rooted in system thinking.

Here we will concentrate upon alphas only and not touching activity spaces and other Essence language elements for kernel (discipline) modeling.

## Engineering solution area of concern alphas

### Differences of software and systems engineering

We cannot use OMG Essence for hardware systems "as is" due to incompatibility its Solution area of concern alphas ("requirements" and "software system") with hardware and sociotechnical projects. We will change these two alphas in the Kernel but will not touch the customer and endeavor area of concern alphas.

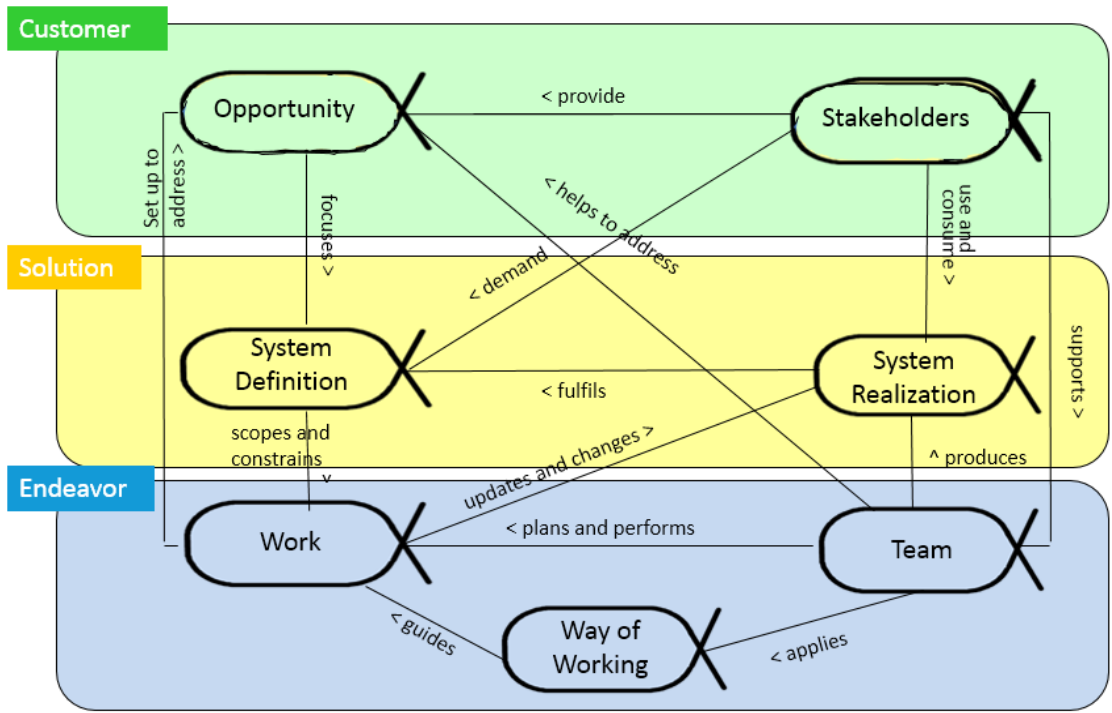

Figure 1: Systems Engineering 7 Kernel alphas

We believe that it is possible have software engineering as a specialization and extension of systems engineering but not vice versa. Systems engineering today is based both on traditional engineering disciplines as well as software engineering.. Thus we need to find most common foundation for all kinds of specialty engineering – including bioengineering, enterprise engineering, control systems engineering, etc. Therefore we need to formulate alphas based not only on the concepts used in software projects. We should generalize Essence based upon system thinking and the usage of a contemporary ontology that permits root system of interest in a physical world that is not restricted by the computer as a media. We will modify Essence Kernel within engineering solution area of concerns and fully preserve Essence Language.

Also we will limit themselves in systems engineering system-of-interest types to a classical software intensive and maximum network intensive systems in a raw of increasing complexity, cumulative ambiguity, "lack of control" that is mecanical and electrical elements; electronic, isolated islands of software; software intensive; network intensive; enterprise, organizational governance (decentralized); system of systems (INCOSE, 2014). Our Essence Kernel modification is not valid to socio-technical systems-of-interest. Nevertheless we still should pay attention to addressing of enabling system that usually is enterprise type but will not change Essence for the sake of it.

**V-diagram**

In (hardware) systems engineering you can easily distinguish alphas of system definition and system realization. You can "kick" or "move" system realization due to its presence in physical world but you cannot do the same with the system definition: pure information as "an abstract object" is not "kickable" or "movable" (while information work products with descriptions are because it is media that is present in the physical world). System realization is present in physical world and thus has spatio-temporal extent (4D: width, height, depth and longevity of its existence in time), system definition provides information about a system and has no such an extent. System definition is "class", i.e. "abstract" or "ideal" object. On the other hand, information work products with descriptions are real world physical objects: paper documents, databases on disks and RAMs.

The same is applicable to software as well. You can think of software (system) realization as a physical object that has spatio-temporal extent. A computer running program code is a kind of physical experiment: running code sets on computer hardware in the starting state of an experiment and after a needed period of code execution you can measure results of that physical experiment (computer hardware state will be changed during program execution until the computation has ended). In software engineering, we can speak about the creation of a software definition as the "development time", software realization time as the "application launch time" and software operations/running as the "run time". However, source code even in machine language (compiled), is not a software realization until it is running in a computer. Thus, one can read a V-diagram for software as consisting of "system definition" and "system realization" branches, as in the case of hardware. At the bottom part of the V-diagram "bits" of software-as-description-of-a-system (e.g. 3D models of hardware or source code of software-as-in-software-engineering) become "atoms", something that have spatio-temporal extent.

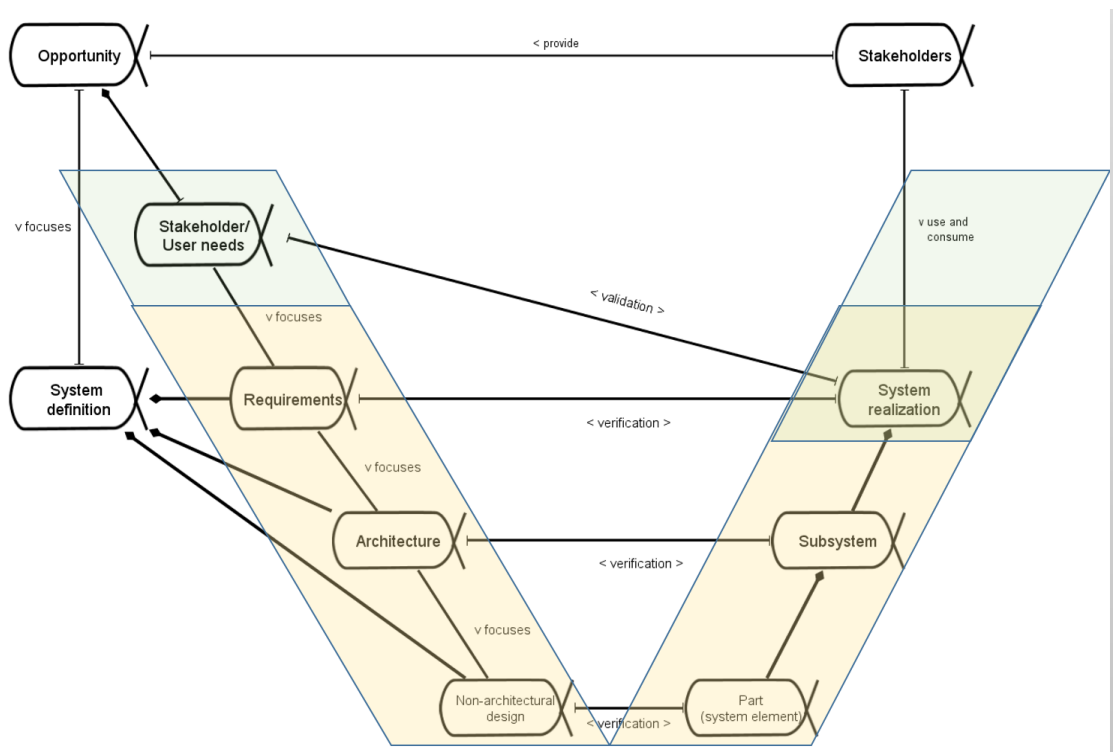

Figure 2: V-diagram expressed in Essence Language

A main heuristic of systems engineering concerns redistributing work towards system definition to eliminate errors during the early stages of development:

abstract bits of data even residing on physical computer media is usually easier to manipulate, check and debug than shaped zillions of atoms in the massive "real thing".

Alpha diagrams are great to show the difference between validation and verification, because these are activities that deal with different systems in overall system hierarchy levels. Engineers cannot apply Essence correctly if they do not know their system (of interest) and their stakeholder/user system as shown in Figures 3 and 4.

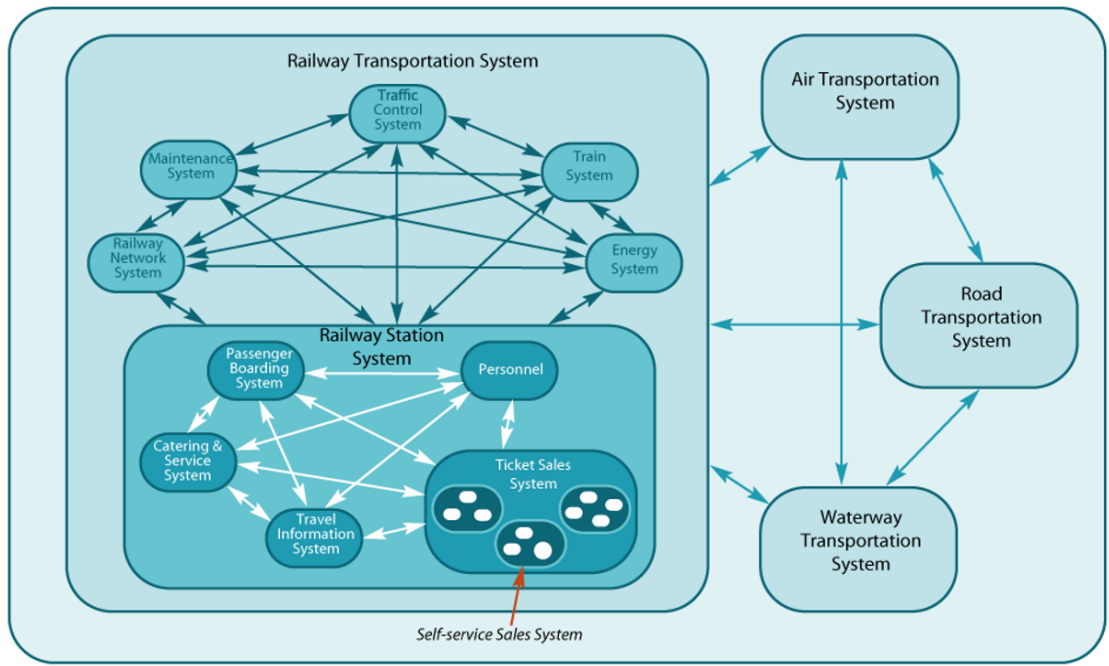

Figure 3. Overall system hierarchy levels as in (Leidraadse, 2008)

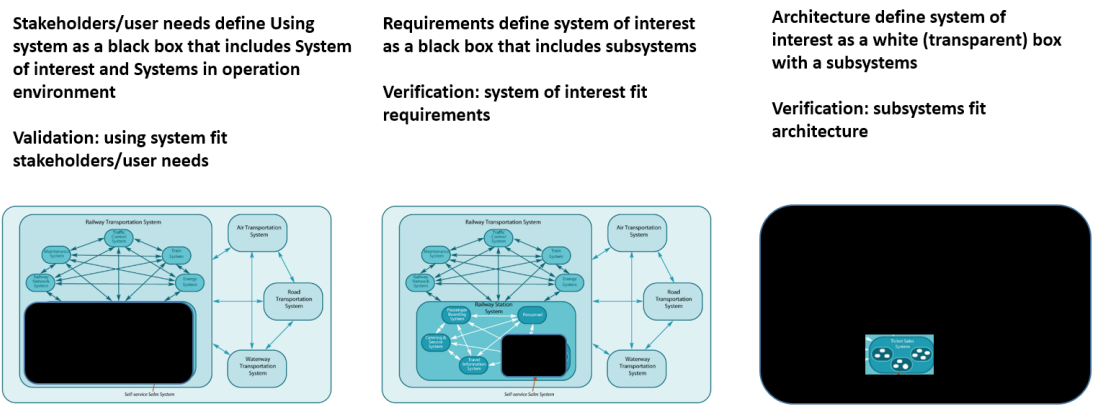

Figure 4: User needs, requirements and architecture in relation to system hierarchy levels.

The use of Kernel alpha diagram and suggested sub-alphas help to better explain requirement engineering practices that should define two different systems and validation and verification practices that should test if these two system Realizations fit to their definitions. Validation is often missing due to inability of engineers to separate Stakeholders/User needs (sub-alpha of Opportunity alpha, clients area of concern) and Requirements (sub-alpha of System Definition, engineering solution area of concern) as related to different systems. Both verification and validation testing including System realization as a project system of interest. In verification they should measure performance of system of interest (i.e. System realization) under test. In validation they should measure performance of using system (with System realization operation as a subsystem of a using system).

This presence of two systems (using and of interest) lead to a necessity of modeling (defining) and even actual creation (realizing) not only instances of system of interest Realization but also instances of using system Realization. It is lead to well known V&V practices like model-in-the-loop, software-in-the-loop, hardware-in-the-loop, etc..

**Generalization of ISO/IEC/IEEE 42010: definitions and descriptions**

The V-diagram reflects multiple systems engineering heuristics and is very general. It is about discipline (alphas) not the technology of enabling system (work products) that usually differ from one project to another, from one engineering enterprise (enabling system) to another. Thus we need to differentiate between the system definition alpha as an ideal-object-from-discipline and a corresponding set of work products.

There are not so many engineering standards that provide a distinction between alphas and work products. ISO/IEC/IEEE 42010 is one of them and it devoted to architectural description. We will map architecture-related ontology of ISO/IEC/IEEE 42010 to the Essence Language and the Kernel systems engineering ontology.

Here is mapping of ISO/IEC/IEEE 42010 to the Language of Essence:

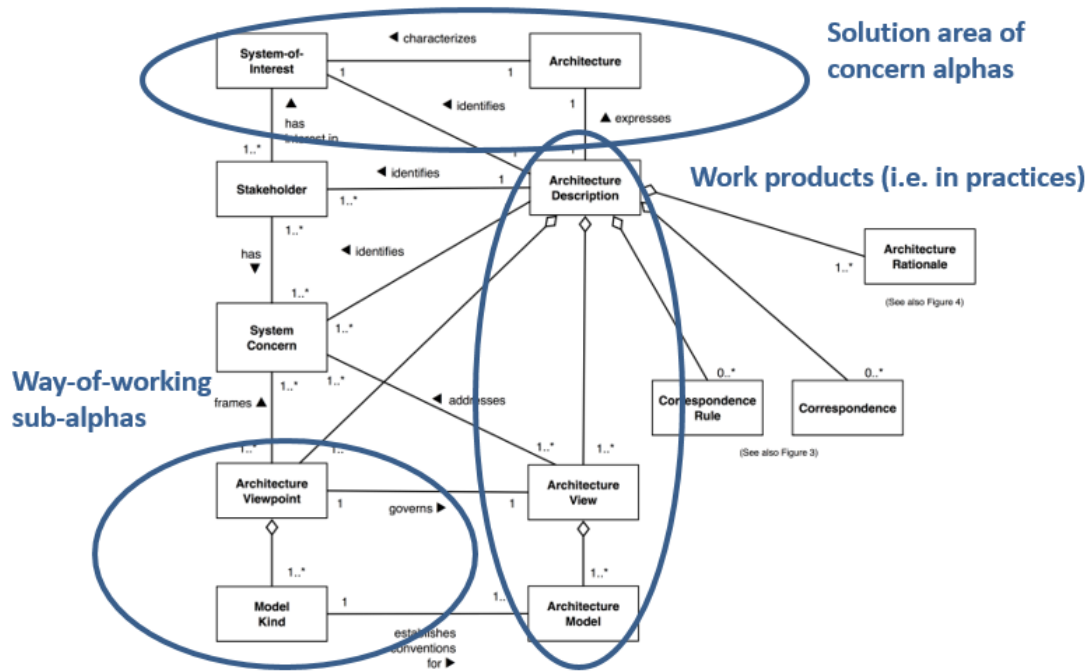

Figure 5: ISO/IEC/IEEE 42010 architectural description concepts mapping to Systems Engineering Essence Kernel

Architecture is an alpha, but architecture description is a work product (ISO/IEC/IEEE 42010 uses the term "express" for their relation, OMG Essence uses the word "evidence" for the same relation).

However, there are additional parts of the system definition in addition to architecture. A system definition is composed from requirements and design. Design is composed from architecture and non-architectural part of design (Clements et al., 2010). You can easily add more (e.g. manufacturing definition like programs for 3D printers, verification definition like tests, etc.). Requirements define the system as a "black box" unlike architecture that defines a system as a "white box", but it is still a part of system definition. Requirements as well as design (architecture and non-architectural) are sub-alphas for system definition. Requirements specifications, standards with requirements stated, product data sheets, etc. are all part of a system description (work products) "express" or "evidence" system definition (alpha).

We can now provide a partial ontology for system definition and description that is compatible with the spirit of OMG Essence, ISO/IEC/IEEE 42010 and the V-diagram together:

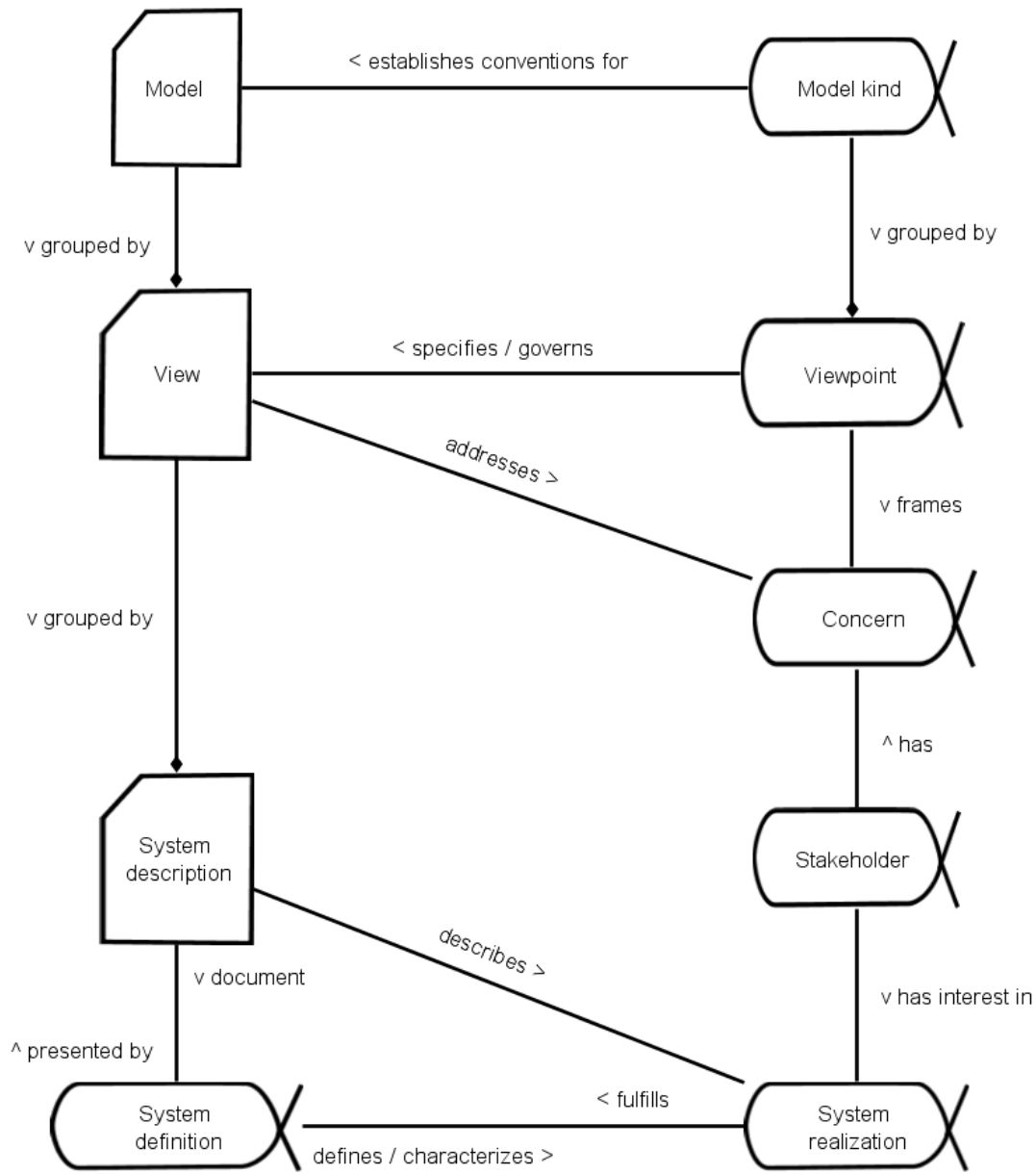

Figure 6: System description related concepts generalized after ISO/IEC/IEEE 42010 architecture description concepts and expressed in Essence Language.

First, every *system realization* (system in reality, system in physical world) has its own *system definition* alpha that presented/expressed/evidenced by *system description* work product (a group of *views* that group their specific models). Second, every *view* of this system description is specified by a *viewpoint* that is in essence (pun intended) way-of-description alpha. The Way-of-description (viewpoint) alpha is sub-alpha of Kernel's way-of-working alpha.

The idea of a system definition alpha supports the idea of alpha as something that exists independent of the degree of documentation of an actual system, like architecture exists for a system even when not documented in an architecture description according to ISO/IEC/IEEE 42010. Likewise requirements (requirements definition) as an alpha exists independently of the degree of documentation of it in a requirements description.

In ISO/IEC/IEEE 42010 we have important note about architecture that applicable also to overall system definition: "The phrase "concepts or properties" is used in the definition [of a term "Architecture"] to allow two differing philosophies to use the Standard without prejudice. These two philosophies are: Architecture as Concept: architecture (of a system) is a conception of a system in one's mind; and Architecture as Property: architecture (of a system) is a property of that system". Thus overall system definition also can be a conception of a system in one's mind (i.e. tacit knowledge); and a property of a system (that can be expressed in a system description).

We can use this diagram to provide designations for system definition work products. According to this diagram a system definition always defines system realization, even on sub-system and system element level. Therefore, if we have solid designation scheme for system realization we could use it as a basis for system description (documents). IEC 81346-1 (IEC 81346, 2009) provides a name for every system in a system breakdown, providing a solid system designation rule set. IEC 61355 (IEC 61355, 2008) provides a way of naming system descriptions (documents) with amending of system designator with a document class. It is very helpful to understand that nothing can be documented that is not related to some system in a system hierarchy and that all the descriptions can be only about systems. Certainly, we should think about amending this designation-related standard set with other elements from a diagram, e.g. providing a means to the designation of a viewpoints (designations for meta-models, method of description etc.), stakeholders, etc.

This diagram also is a good demonstration of connection between science and engineering. Science provides theories that are available as a library (i.e. reusable, already known) of viewpoints, engineering using these viewpoints to synthesize system descriptions. In OMG Essence viewpoint belongs to the Way-of-Working alpha (viewpoint is system description practice).

### System realization alpha

Chronologically a system definition alpha should be treated first in the engineering life cycle. However, the system of interest is actually the realized system. Stakeholders need a system realization alpha to address opportunity alpha. We should understand first what the system realization would become with an adequate system definition. The same cause is behind heuristics that business processes should be discovered "from the output to inputs" while executed "from the inputs to output" order. In the V-diagram development process we will describe here "from the system realization to definition" while executing in the order "from the system definition to realization".

### Multiple views and viewpoints - Separation of concerns

A system approach reveals the need for multiple viewpoints. Every viewpoint generates (in the sense of "generative grammars" (Chomsky, 1956)) – multiple views that describe (evidence, express) the state of the system definition alpha. Multiplicity of descriptions is the main way of coping with complexity of systems. Separation of concerns that is famous in software engineering addresses the same aspect (Dijkstra, 1982).

The Engineering ontology standard ISO 15926 suggests a way of thinking about multiple descriptions that have borrowed from 4D extensionalism philosophy (Partridge, 2005). If any two descriptions evidence their objects as existing in the same spatial-temporal extent, it means that both of them are descriptions of the very same 4D object. 4D extensionalism provides a way to establish a correspondence rules between different views.

An example of correspondence can be modeling of the equivalence of pump on P&ID diagram (FunctionalPhysicalObject, have a functional designation on a drawings) and pump in place (InanimatePhysicalObject, having a serial number). Both refer to the same spatio-temporal extent, actual pump that is installed according to drawings and has a vendor serial number.

One more feature of the ISO 15926 is the ClassOfClass type that permits neglecting a main or preferred classificator from some exclusive view/viewpoint of particular discipline. There is no such a thing like "main" view and "main" viewpoint. This is reflected also in IEC 81346, ISO/IEC/IEEE 42010 and reflects the system thinking that is interdisciplinary and has no lead view/viewpoint pair.

Essence also supports this multi-viewpoint/separation of concerns principle. But 4D extensionalism that is about hardware (including information media, but not ideas/information) provides additional rigor to system thinking and supports integrity in the composition of multiple models in system views to an overall system description.

**Components, modules, allocations**

How many types of structures exist in systems? (Clements et al., 2010) tells us that there are at least three "architectural styles" that correspond to three types of elements: components and connectors, modules and interfaces, allocations. Certainly there are many more than these three, and usually we see mainly hybrids, not pure system descriptions that depict only one type of system element. An important note in (Clements et al., 2010) is that architecture should include minimally one description (view) from each of the style type (viewpoint) indicating that less than three views is not yet a viable architecture.

IEC 81346 indicates the same thing but with wording from construction industry speech community: there are functional aspects (components in (Clements et al., 2010)), product aspects (modules in (Clements et al., 2010)) and location aspects (allocations in (Clements et al., 2010)). Computer memory or a warehouse location is still a location in space!

In the aerospace systems engineering speech community (Kossiakoff, 2011) identifies system realization objects as functional elements (components in (Clements et al., 2010)) and components (modules in (Clements et al., 2010)). Pay attention: this is one systems engineering semantic community, but in the language community we have strictly opposite words usage: what (Kossiakoff, 2011) calls "component" is module in (Clements et al., 2010) and vice versa.

ISO 15926 provides the same distinction between FunctionalPhysicalObjects (components, functional elements) and InanimatePhysicalObjects (modules, products). FunctionalPhysicalObjects in ISO 15926 is "physical", you can "eliminate" them due to their existence in the physical world. It is not a "logical" abstraction while it definitely is an element of functional decomposition of a system.

Usual systems engineering notes that synthesis of a system architecture suggest that it appears in mutual adjustment of the logical architecture (functional decomposition) and physical architecture (module/product decomposition).

It appears that the consensus concerning system realization is that it should be a component, a module and a location in space. These three types of objects appear as the very same object (system realization) according to 4D extensionalism. Certainly, the system of interest (aka system realization) is not only a component a module and a location in space but is a subcomponent, submodule, sublocation in space for the upper level system and consists of subcomponents, submodules, sublocations in space as subsystems or system elements.

Components, modules, allocations all are sub-alphas of the system realization alpha, it is a system that consists (part-whole relation) of components, modules, (al)locations, not system definition. System definition only defines in its description work products all these components, modules, allocations and their relations (correspondence).

**System breakdowns and designations**

A process plant or a submarine can have more than 3-10 million individual parts. There should be way of naming (designation) for all of these parts, each of which can be defined as a separate system in the overall system of interest, i.e. can be component, module, or allocation. IEC 81346 indicates that three main system structure types provide minimally three kinds of designators (component/functional element, module/product, allocation) and these designators should be structured according to a system hierarchy ("true" system breakdown), reflecting part-whole relation between systems and subsystems in the system realization. There is no single "system structure", there are at least three of them.

IEC 81346 suggests that a system designation should reflect the multiview/multiaspect character of a system and include at least partial mention of multiple system realization sub-alphas (e.g. =F1 / -12-N4-DN18 / +M13 identifies a system that is component F1, module DN18 that is sub-module of module N4 that is in its turn sub-module of module 12 and resides at the M13 location. Not all of the aspect designators must be unambiguous, but at least one of them should be).

Interestingly enough is that in software engineering designators of different aspects rarely can be seen together and usually have no indications of the view/aspect to which they belong. E.g. names of components/functional elements usually remain in architecture diagrams and the sources consist mainly of module names/designators, and make/configuration files devoted to allocations.

**System realization alpha states and their checkpoints**

We now consider a variant of the system realization alpha states and their checkpoints. It roughly follows the system realization part of the V-diagram. It is more "cascade-oriented" than agile as in the original "software system" alpha of OMG Essence. Nevertheless, it is a typical of software intensive hardware and not a pure software system.

According to ISO 15926 every system realization state is a TemporalWholePart of a system realization (that is the whole system for a specified period of time). It is very convenient that multiple different types of relations that are needed to describe changes in time (including e.g. change of a pump that failed validation testing) is reduced mainly to different kinds of PartOf relation. This is a power of 4D extensionalism for engineering.

| System realization states | System realization state checkpoints |
|---|---|
| Raw materials | Raw materials for system realization are available and allow manufacturing of the parts with required properties.<br>• Facilities for manufacturing parts from the raw materials are available.<br>• Parts production and logistic schedule has been agreed.<br>• Parts manufacturing facilities are ready to start. |
| Parts | Parts have been produced and are ready for integration.<br>• Parts of the system have been produced and/or purchased and checked.<br>• Integration schedule has been agreed.<br>• Integration facilities are ready to start. |
| Demonstrable | The system has been assembled from the parts and is ready for testing.<br>• Some functions of the system can be exercised and key characteristics can be measured.<br>• Key system characteristics have been demonstrated.<br>• Critical interfaces have been demonstrated.<br>• The integration with other existing systems has been demonstrated.<br>• The relevant stakeholders agree that system has been tested. |
| Ready | The system (as a whole) has been accepted for deployment in a live environment.<br>• The functionality of the system has been tested.<br>• Level of defects is acceptable for the stakeholders.<br>• Setup and other user documentation is available.<br>• The stakeholder representatives accept the system as fit-for-purpose. |

|  | • Configuration of the system to be handed over to the stakeholders is known. The stakeholder representatives plan to make the system operational. The system is fully supported to the agreed service levels. |
|---|---|
| Operational | The system is in use in a live environment. <br>• The system has been made available to the stakeholders that intended to use it. <br>• At least one example of the system is fully operational. <br>• The system is fully supported to the agreed service levels. |
| Retired | The realized system is no longer supported and disposed and/or recycled. <br>• The system realization has been replaced or discontinued. <br>• The system is no longer supported. <br>• There are no "official" stakeholders who still use the system. <br>• Updates/ modifications to the system will no longer be produced. <br>• All material components of the system are re-used or have been properly disposed. |

Table 1: System realization alpha states and their checkpoints

These states and the checkpoints are far from being a generalization for all kinds of system realizations. If we can think about generalization of system definition, it hardly suggests some kind of general realization framework. While definition is mainly about thinking and describing and therefore rather homogeneous as a domain, realization is drastically varying in its nature and belongs to a variety of domains. It is not only about wording (variety in concept names), but is about human action variety (variety of realization-related concepts that stem from variety of world objects and related activities).

We have a relatively small number of "designing world" concepts that we need to generalize for system definition. But we have drastically different "production world" concepts. E.g. to "design" of a plant, spaceship, software, building will be well understood with one term. But "manufacture" all this will be strange: we need different concepts to describe the process – "manufacture", "assemble", "build", "construct", and even "deploy" in software. If we want to provide a general checkpoint list for the system realization alpha it is better that we use "realization" (literally "transfer to reality") as the term rather than trying to specify it more precisely. May be we need to specialize this alpha for different industry domains to better reflect this variability in system realization alpha state sequences and corresponding checklists.

You can consider what a system realization alpha state can be in the case of a pure software engineering project (if any exist), cyber-physical system of interest (e.g. anthropomorphic robot) or a loosely commanded system of system (e.g. socio-technical system, every human in this system is a rather autonomous system and their realization as a system modules is usually "placement" and "training" with a heavy dose of somebody's leadership). It is hard to generalize all these in one-size-fit-all system realization alpha states and checkpoints for every state.

### System definition alpha

The System definition alpha contains sub-ordinated alphas such as requirements, architecture and non-architectural design. All of sub-ordinated alphas of system definition exist and have their states independent of the degree of their documentation.

### Requirements

A specific systems engineering meaning of requirements, is a "black box" (without knowing about subsystem decomposition) definition of a system. Functional requirements define a system of interest as a component. Interface requirements define the system of interest as a module. There may also be system allocation requirements and many others according to different views/aspects and type of system elements structure in a particular view.

### Architecture and non-architectural design

A White (transparent, glass) box system definition is a design that includes "architecture and non-architectural design" (Clements et al., 2010). Architecture we understand is in full compliance with ISO/IEC/IEEE 42010.

Non-architectural design is not directly part of systems engineering concerns. But it is very important part of overall system definition. Architecture is (subjectively from a system engineering perspective) is a set of most important engineering decisions that when changed lead to an almost complete redesign of a system of interest. Non-architectural design is remaining set of myriad other factors that are not so important engineering decisions that when changed do not lead to major redesign work. Non-architectural design usually is large in volume in comparison with architectural part of a design.

Conditionally we consider programs for numeric controlled machines as descriptions (work products) expressing non-architectural design alpha. Manufacturers would be willing to have separate "production technology" alpha as a sub-alpha of system definition alpha. Still this is different from a software process where "production/deployment" is very easy and well automated.

### System definition alpha states and their checkpoints

We now consider a variant of system definition alpha states and their checkpoints. It is roughly following the system definition part of V-diagram. Unlike being special for each kind of engineering system realization alpha states and checkpoints, system definition alpha state and checkpoints can be easy generalized for different (system, hardware, software) engineering. The result of an execution of activities from a system definition activity space is a structured set of system descriptions: documents and databases that are very diverse by content and methods of their creation, management and usage but somewhat similar in form and function. But we differentiate from the original Essence where only requirements are used as system definition but architecture is subordinate to software system alpha.

| **System definition state** | **System definition state checkpoints** |
|---|---|
| Conceived | It is clear how the system will be defined.<br>• It is clear what success is for the new system.<br>• Viewpoints are agreed upon.<br>• The approach to concord descriptions among the stakeholders has been agreed.<br>• The description of change management mechanisms have been agreed. |
| Consistent | Consistent System definition has been created.<br>• Descriptions are documented and available for the team and stakeholders.<br>• The origin of the description is clear.<br>• Descriptions are examined.<br>• Contradictory descriptions have been identified and are dealt with.<br>• The team understands descriptions and agrees to implement them.<br>• The system implementing the descriptions is accepted by the stakeholders as worth realizing. |
| Used for Production | System definition is used for system production.<br>• Enough of the descriptions are ready for starting system realization.<br>• Realization technologies have been defined.<br>• Those responsible for system realization part of the team acknowledges that available descriptions are sufficient to realize the system.<br>• Issues occurring during system realization lead to the re-work and actualization of the system definition. |
| Used for Verification | System definition is used for testing.<br>• There are no missed parts of the system definition that make testing impossible. |

| | • Tests, success criteria and test methods have been defined.<br>• Stakeholders agree with test scope. |
|---|---|
| Used for Operation | System definition is used by stakeholders for operation.<br><br>• System definition is used for gathering information about state of the operational system realization.<br>• System definition is within the information about the state of the operational system and is used for making decisions about maintenance, repair, and modernization. |
| Used for Disposal | System definition is used for system disposal.<br><br>• System definition is used for making decisions about system disposal or operation extension.<br>• System definition shows absence of undesirable consequences (e.g. environment pollution) through system disposal.<br>• System definition is used for planning and performing disposal or recycling of the system realization. |

Table 2: System definition alpha states and their checkpoints

### Endeavor definition

ISO/IEC/IEEE 15288 and OMG Essence indicate that contemporary systems engineering is not only about the engineering solutions area of concern. System engineers should also keep in mind and actively participate in progressing other domains (client and endeavor) kernel alphas.

#### Enterprise engineering

Enterprise engineering is increasingly important application of systems engineering not only to system of interest but also to enabling systems (in ISO/IEC/IEEE 15288 terms). There should be significant harmonization of the enterprise engineering discipline and practices with systems engineering methodological frameworks and this is no less important than harmonizing system engineering with software engineering.

Essence provides an idea of how to apply systems engineering more directly to enterprise systems. You should keep in mind that enterprise development projects can suggest work to be done on all the possible system of interest development projects, but we schematically simplify this as ISO/IEC/IEEE 15288 did: consider endeavor context only of one system of interest development project. You can call it an endeavor development project (where system of interest is the endeavor).

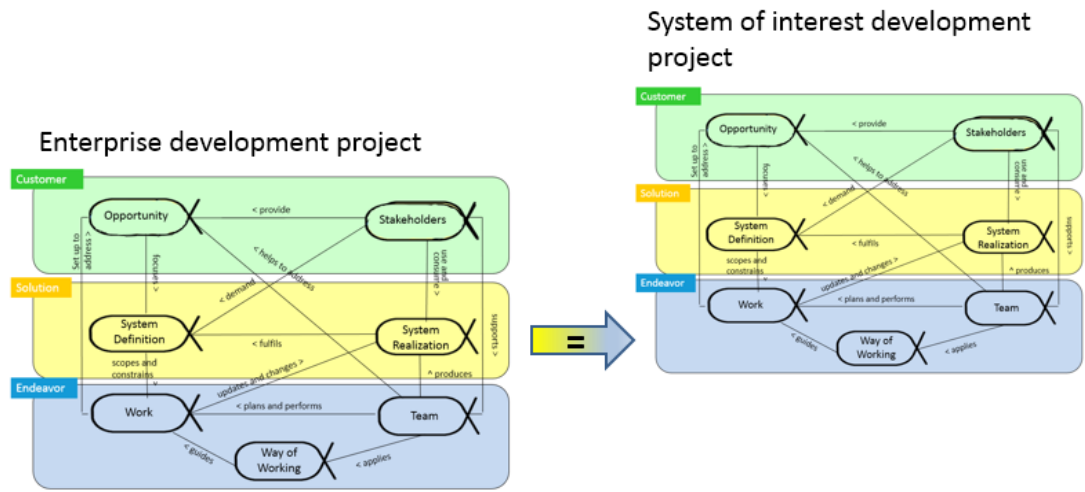

Figure 7: Enterprise development as endeavor development

While there are multiple differences in terminology ("letter") there are no significant differences in the concepts ("spirit"), same distinction of the semantic and the speech community. E.g. instead of requirements sub-alpha of system definition enterprise engineers indicate strategy and motivation (OMG BMM, 2010; ArchiMate, 2014), where system architecture becomes enterprise architecture, validation and verification become corporate governance and measurements.

There is a close bond between enterprise/endeavor development projects and traditional systems engineering projects, part of this field is treated by the systems engineering management sub-discipline. E.g. Conway's law (Conway, 1968) indicates the close bounds between system of interest structure and enabling system structure. That means that when you significantly change the architecture of a system of interest you should change in turn the architectures of the enabling systems, i.e. architecture of an endeavor/enterprise/extended enterprise.

Still you should be very careful in application of systems engineering Essence Kernel to such system of interest as an enterprise and/or system of system. This must needs further explorations.

**Work description approaches (viewpoint types): activity-based, product-based, communication-based**

How should an endeavor/enterprise be described? What structures do you have in an enterprise? The Essence answer is that a Team performs Work according to a Way of Working.

There are three basic viewpoints that stem from such a definition: practice-based (fits best to answer stakeholder's questions about the Way-of-working), process-based (project management related to Work alpha), team-based (related to communications, authority and responsibility of the Team alpha).

The same practice-based, process-based and team-based in (Cordys, 2008) calls respectively artifact-based (due to importance of work products aka artifacts to practices), activity-based (process as sequence of activities), communication-based (one of important things about a Team is the communication about Work division for a given Team, authorities and responsibilities of team members). The same three viewpoints in (Wang et al., 2005) are called respectively information-based (in most non-manufacturing companies work-products mainly information objects), process-based and organization based.

Essence favors a practice-based approach in the (enterprise/endeavor) system definition with accent on alphas and work products and is therefore closer to adaptive case management and issue tracking than classical project management (Swensson, 2010), ISO/IEC/IEEE 15288 favors cascade and project management (process approach) that is more appropriate to huge hardware plant related projects like aerospace and defense, nuclear plants and others that is difficult to imagine as being "agile". Team-based management methodologies and definitions are almost absent in software and systems engineering. But knowing the importance of using all needed viewpoints indicates the need to proactively seek team-based system definitions and team-based practices. The DEMO methodology is an example of such an approach (Dietz, 2006).

One thing that a systems engineer can inform an enterprise engineering colleague about is "you need'em all!". Do not consider only using an issue tracker without project management tools, do not consider the option of only using project management tools and ignore change management. Do not forget about structuring of a Team! And be sure that all these are about the same object, the same endeavor and the same (organization) system realization.

## Conclusion

This approach towards establishing a Systems Engineering Essence framework was actively discussed at a meeting of the INCOSE Russian chapter and SEMAT Russian chapter. In the Fifth Systems Engineering Challenges Workshop 2014, there were six reports of using this framework in actual engineering projects. But still there is more success with applying this framework usage in software-intensive and information system engineering projects than in classical non-computer hardware projects.

In 2013 and 2014 years, a PhysTech 4 credit course to introduced systems engineering propaedeutic to mainly physics-educated postgraduates occurred. It was provided at Ural Federal University. It appears that in comparison with multiple other options (e.g. classical "Foundation of Systems Engineering" course that was given in PhysTech 2012) that the course students demonstrated more understanding of what was happening in their educational engineering projects. In their reports, most of all the students mentioned the convenience of thinking and communication in terms of alphas while each having extremely diverse systems engineering endeavors (classical systems engineering, software engineering, enterprise engineering) with particular tools, work products and terminology preferences. Another thing was reported that students gave more attention to non-engineering (clients and endeavor) domains of interest that provided them with understanding of a proper context for their engineering solutions and more rapid adaptation to their work (PhysTech and Ural Federal

University both have a tradition of using real world engineering projects as a base of innovation-minded physics-intensive engineering enterprises).

The Systems Engineering Essence framework also was presented to the nuclear industry in a set of tutorials for specialty engineers to give them a grasp of systems engineering thinking. This also was successful. A main remark was about the importance of simultaneous usage of different views to a system and attention not only to system of interest in operational environment but also to enabling systems. Engineers that have engineering management tasks in their respective enterprises provided feedback that after tutorial they better understand what they are doing and why it requires more knowledge to deal with endeavor area of concern alphas (every alpha has devoted disciplines to work with them).

But it was difficult when the Systems Engineering framework was introduced at one of high capacity industrial chillers design bureau since a practice-oriented viewpoint severely conflicted with the project-management process-based approach that was the custom for not only engineers but to managers of that particular enterprise. In waterfall environments, attention is usually not focused on alpha states but activities/activity spaces. We can expect that this can be the main difficulty in applying a Systems Engineering Essence in typical systems engineering projects with a low level of agility. We need more attention on harmonizing and the simultaneous use of practice-based viewpoints to endeavors corresponding to the alpha and product states Essence cards, case management and issue tracking tools with a process-based viewpoint supported by project management and BPM tools.

Overall, comparing and harmonizing multiple software engineering and systems engineering standards is possible based on a system perspective and distinguishing Language, Disciplines (Kernel and its tweaking and amendment) and Technologies (practices that implement Disciplines with certain work products and tools). The suggested preliminary draft of a Systems Engineering Essence has already proved to be helpful and will be maturing in years to come.

## Acknowledgements

Thanks for INCOSE Russian chapter members for multiple discussions on the topic. Special thanks to Bud Lawson for help in structuring and editing.

Thanks to Bud Lawson, Don O'Neill, Ilia Bider and Rich Hilliard for valuable comments on an early draft of this paper.

Sowa, J. (2012), Criteria for evaluating ontologies at different levels, http://ontolog.cim3.net/forum/ontology-summit/2012-12/msg00131.html

Swensson, K. (2010), Mastering the Unpredictable: How Adaptive Case Management Will Revolutionize the Way That Knowledge Workers Get Things Done (Landmark Books), Meghan-Kiffer Press

Wang, J., et al. (2005), A framework for Document-Driven Workflow System, W.M.P. van der Aalst et al. (Eds.): BPM 2005, LNCS 3649, pp. 285–301, Springer-Verlag Berlin Heidelberg

## Appendix

Most of the systems engineering standards contribute to one or another part of an overall system perspective. What we tried to show in this text is that a more complete system landscape appears from multiple parts scattered in a set of standards. Below is short description of mentioned standards and standard-like related publications (in alphabetical order).

*BKCASE, Body of Knowledge and Curriculum to Advance Systems Engineering (2014), http://www.bkcase.org/* Two publications (Body of Knowledge and Curriculum maintained independently) that capture systems engineering knowledge and knowledge about systems engineering education. This is joint effort of INCOSE, CS IEEE, Stevens Institute of Technology. BKCASE is more like recommendation standard for education institutes (unlike SEH, INCOSE Systems Engineering Handbook that is used in certification of systems engineering professionals). In future versions of the BoK it is planed to provide harmonization with SEH, ISO/IEC/IEEE 15288 and PMI PM BoK (project management body of knowledge standard of Project Management Institute). Also BKCASE editors plan to get help from the International Society for the Systems Sciences (ISSS) to better understand and reflect in BKCASE the relationships between systems science and systems thinking as applied to engineered systems.

*IEC 61355-1 (2008), Classification and designation of documents for plants, systems and equipment – Part 1: Rules and classification tables.*

This standard suggest classification schema of documents as well as documents kinds and suggest designations for them. The standard suggests a document kind classification code (DCC) with prefix & and three letter codes for technical area (A- overall management; M-mechanical engineering, normally including process engineering; etc.) classes and subclasses (CA – contractual and nontechnical documents-inquiry, calculation and offer documents). It permits easy designation of hundreds of thousands documents that occur in a major engineering effort such as chemical or power plant design, construction and operation. One of the ideas is that any document describes "something" and in an engineering "something described" usually is a system, thus document designations consist of described system designations ("tag", defined in IEC 81346) and followed after & prefix document designation.

*IEC 81346 (2009), Industrial systems, installations and equipment and industrial products -- Structuring principles and reference designations -- Part 1: Basic rules.*

This standard jointly defines rules for system structuring and a designation for a resulting system structure. The standard suggests that number of levels in a system is undefined (most of previous standards prescribe exact number of system level – e.g. system – element or segment – subsystem – assembly – subassembly – component – part), especially in supply chains. What is a component in one engineering project is a full-fledge system with multiple levels of hierarchy in another engineering project. Thus, we need free ourselves from fixed-length designators and simultaneously discuss system structure principles and principles of designations. In addition, this standard suggests simultaneous usage of designators for different aspects of a system and declares that every aspect has its own system breakdown structure.

*ISO/IEC/IEEE 12207 (2008) Systems and software engineering - Software life cycle processes and ISO/IEC/IEEE 15288 (2008) Systems and software engineering - System life cycle processes.*

This pair of standards is currently in a joint harmonization process. While they currently define different set of practices (processes) that are devoted to software and systems engineering respectfully, both of them used the same ISO/IEC TR 24774 guidelines for process description. Both standards are rather new for engineering: both take system approach to system of interest. They define enabling systems/project/enterprise (engineering management) part and not only technical practices/processes of engineering; both standards cover the full life cycle and both can be applied at any level of system decomposition.

*ISO 15926-2 (2003), Industrial automation systems and integration -- Integration of life-cycle data for process plants including oil and gas production facilities -- Part 2: Data model.*

This standard is about engineering information systems federation to provide a consistent system description in all the stages of system life cycle. It suggests common and general for any particular specialty engineering data model that is neutral in relation to software vendors of CAD/CAE/PLM information systems. Unlike ISO 10303 (STEP) standards family that failed to provide one common data model for all engineering domains. ISO 15926 suggests a semantic/logical/fact-oriented data model (not object-oriented) because "what is object in one engineering information system data model is attribute in another and vice versa". This is unlike many similar contemporary initiatives (BIM, building information model; CIM, common information model; contemporary enterprise level standards for engineering master data management) that all are object-oriented like in object-oriented programming. Moreover, this standard defines an engineering ontology that is founded on strong philosophical foundation and aims at description changes of system definition and system realization during life cycle of a complex and large engineering project. Certainly, this ontology supports a system approach and engineering thinking about systems and their changes and properties, not IT-thinking in terms of "data elements" and "data fields and pointers/keys": it is about engineering systems (system elements and how to process and compose them), not computer-focused data presentation (data elements and how to read/write and combine them).

*ISO/IEC 24744 (2007), Software Engineering -- Metamodel for Development Methodologies.*

This standard is the latest of the first generation of situational method engineering standards. It defines a Language for engineering endeavor description, but not a corresponding Kernel. It has more academic than industrial value due to the absence of proper software tools that support modeling of practices with it's metamodel. Most prominent of its innovations was the usage of "clabjects" – special entities that have the property of classes and objects simultaneously. This was important for properly modeling of methodology realm classes and endeavor realm instances. Early standards were not capable of description of methodology and endeavor realms entity relationships.

*ISO/IEC TR 24774 (2010), Systems and software engineering -- Life cycle management -- Guidelines for process description.*

This standard is often confused with *ISO/IEC 24744 (2007), Software Engineering -- Metamodel for Development Methodologies* that have similar digits in it's identity, but it is completely independent. This standard suggests naming conventions and four levels of decomposition for "processes"(activity-related part of practices) for ISO/IEC/IEEE 15288, ISO/IEC/IEEE 12207 and several other "process framework" standards. From situational method engineering viewpoint this standard looks like a small part of not too advanced (e.g. due to fixed level of decomposition, ignoring work products, relation of endeavor and methodology realm, etc.) and a language for practice description.

*ISO/IEC/IEEE 42010 (2011) Systems and software engineering - Architecture description.*

This is the first standard that formalized the notion of architecture in relationship to software systems, but it has been accepted in the systems engineering community as well and even in the enterprise architecture community. This standard is a base of multiple architectural frameworks and languages standards that follow the suggested terminology and ontology of architecture and architectural description, view and viewpoint distinction, architectural models as main constituents of architectural description and correspondence rules that join models of different views. This standard seriously takes a system approach into account: system, stakeholder, stakeholder concerns are drive all architecting and provide multiple views, not "one view-fit-all" like in reductionist approaches.

*OMG BMM (2014), Business Motivation Model*

This specification is based on SWOT (strength, weaknesses, opportunities, threats) analysis framework that is often used in structured business planning effort: it defines a conceptual structure for business plans. It is usually used with OMG SBVR and OMG BPMN to provide an object-oriented way of defining business architecture in overall enterprise architecture in form convenient to software programmers. It is object-oriented and is compatible with multiple other standards of OMG. Thus, it is less of an enterprise architecture standard but more of a IT-solution architecture standard that should capture organizational governance knowledge for software architects.

*OMG SBVR (2013), Semantics of Business Vocabulary and Rules*

This is ontology standard like ISO 15926, but SBVR is more about enterprise engineering and business rules than systems or software engineering. It is about computer reasoning with business rules. If you have no exact meaning to every term used in those rules, all the reasoning will be meaningless. Therefore, SBVR defines a Language for ontology and rules representation along with some Kernel for enterprise (organizational) ontology. It supposed that SBVR should be used with BMM, to give ontological rigor not only to business rules but also to business plans.

OMG SPEM 2.0 (2008), Software & Systems Process Engineering Metamodel,

Most popular of first generation of situational method engineering standards that provided tool support. This standard is in use in life cycle processes descriptions in projects with heavy "cascade" life cycle models.

*OMG SysML (2012), Systems Modeling Language*

This is "shrinked and expanded" (i.e. extension of a subset) UML that was developed specifically to modeling of requirements and architecture in systems engineering. Along with AADL (Architecture Analysis & Design Language) SysML is popular for architecting activities in systems engineering projects. Despite its name it does not directly support a systems approach in its metamodel.

*SEH (2014), INCOSE Systems Engineering Handbook*

This "Handbook" is an INCOSE publication that captures systems engineering knowledge that in use to train and certify systems engineering professional in INCOSE. Alternative publication is BKCASE, but actually, in professional exams INCOSE utilizes the "Handbook" and not the BKCASE publication.